\documentclass[prd,nopacs,nokeys,nofootinbib]{revtex4}
\usepackage{amsfonts}
\usepackage{amsmath}
\usepackage{amssymb}
\usepackage{bbm}
\usepackage[utf8]{inputenc}
\usepackage[caption=false]{subfig}
\usepackage{tensor}
\usepackage{slashed}
\usepackage[centertableaux]{ytableau}
\usepackage{hyperref}
\usepackage{natbib}
\usepackage{braket,mleftright}
\usepackage{graphicx}
\usepackage{cleveref}
\usepackage{tikz}
\usepackage{booktabs}
\usepackage{tikz} 
\usepackage{tikz-feynman}
\usepackage{tikz}
\usepackage{subcaption} 

\begin{document}

\title{Macroscopic Signatures of Gauge-Mediated Contagion: Deriving Behavioral Shielding from Stochastic Field Theory}

\author{Jos\'e de Jes\'us Bernal-Alvarado}
\email{bernal@ugto.mx}
\affiliation{Physics Engineering Department, Universidad de Guanajuato, M\'{e}xico}

\author{David Delepine}
\email{delepine@ugto.mx}
\affiliation{Physics Department, Universidad de Guanajuato, M\'{e}xico}

\date{\today}

\begin{abstract}
We present a unified theoretical model relating stochastic microscopic epidemic dynamics with macroscopic non-linear population behavior. Utilizing the Doi-Peliti formalism, we model the pathogen as a gauge mediator field coupled to susceptible and infected host populations, and introduce a Reactive Immunity Field capable of spontaneous symmetry breaking. We demonstrate that the naive epidemic vacuum is destabilized by radiative loop corrections via the Coleman-Weinberg mechanism, generating a dynamic herd immunity threshold. By extracting the classical saddle-point limit of the Effective Action, we  derive the macroscopic reaction-diffusion equations governing the host population. We show that integrating out the gauge mediator inherently generates a thermodynamic Free Energy dependent on the square of the susceptible density. This non-linearity produces a macroscopic spatial ``Fear Drift'' proportional to the magnitude of the immunity field, and a cubic shielding penalty in the effective reproductive number ($R_{eff}$). In this work, we establish a  mapping between fundamental field-theoretic mechanisms and specific terms in the macroscopic behavioral equations. We demonstrate that Debye screening is physically executed by the spatial cross-diffusion fluxes driving host evacuation. Simultaneously, vacuum polarization manifests as a non-linear cubic penalty ($-S^3 I$) in the dressed reaction rate that dynamically suppresses the effective reproductive number.
As a validation of our model, we apply the formalism to high-resolution spatiotemporal COVID-19 data from Germany. By treating the infected population as an extensive thermodynamic variable and behavioral shielding as an intensive gauge field, we empirically demonstrate macroscopic bistability and gauge-driven hysteresis. The emergence of a double-well potential and non-Markovian phase space trajectories (``Fear Drift``) confirms that epidemic propagation is governed by topological phase transitions and societal inertia, offering a  departure from memoryless mass-action kinetics and providing novel physical insights into the dissipative costs of delayed public health interventions.
\end{abstract}

\maketitle
\section{Introduction}

Classical epidemiological models, such as the standard Kermack-McKendrick SIR model \cite{kermack1927contribution,anderson1991infectious, murray2002mathematical}, rely heavily on mean-field approximations that assume well-mixed populations and static interaction parameters. While these deterministic models are highly effective for capturing the basic trajectories of moderate pathogens, they consistently fail to predict the severe non-linear spatial dynamics, burst clustering, and behavioral feedback loops observed during modern pandemics involving highly infectious variants \cite{brockmann2006scaling, brockmann2013hidden,gonzalez2008understanding}.

To address these limitations, recent advancements have mapped epidemiological dynamics onto the framework of Quantum Electrodynamics (QED) via the Doi-Peliti second-quantization formalism \cite{doi1976second,peliti1985path,tauber2014critical}. In this paradigm, the pathogen is not treated as a simple reaction rate, but as a continuous gauge mediator field ($\varphi$) that carries the ``epidemic charge'' between localized host populations.

In this work, we extend the QED-inspired SIR model given in ref.\cite{bernalalvarado2026}, introducing a continuous Reactive Immunity Field ($\Phi$) to describe the dynamic, spatial emergence of herd immunity. Rather than treating the epidemic threshold as a static depletion of susceptible hosts, we investigate the stability of the epidemiological vacuum using the Coleman-Weinberg mechanism \cite{coleman-weinberg}. We demonstrate that the radiative loop corrections of the pathogen field induce spontaneous symmetry breaking, effectively granting the pathogen a screening mass and limiting its propagation.

Crucially, we relate this microscopic quantum field theory to observable macroscopic phenomena. By taking the classical saddle-point limit of our model, we derive the  temporal and spatial macroscopic equations governing the host populations. We reveal that the gauge-mediated interaction  alters the spatial transport of the susceptible population, generating non-linear cross-diffusion terms such as ``Infection-Activated Dispersion'' and ``Fear Drift''. Furthermore, we show how these macroscopic behavioral responses dynamically affect the effective reproductive number ($R_{eff}$), explaining the mechanisms behind critical opalescence, premature epidemic burnout, and endemic bistability.

The ultimate test of any physical theory lies in its empirical verification. Therefore, a central contribution of this paper is the macroscopic validation of our QED-inspired formalism using comprehensive spatiotemporal data from the SARS-CoV-2 pandemic in Germany (over 400 districts). Unlike standard compartmental models that require arbitrary parameter forcing to fit real-world data, we demonstrate that the natural inclusion of a  screening mass is giving us a effective reproduction number ($R_{eff}$) much closer to the empirical one that can be obtained from data than the usual SIR models results. Furthermore, by constructing the macroscopic phase space of the epidemic,  topological signatures previously undocumented in classical epidemiology are obtained: spontaneous symmetry breaking leading to endemic bistability, and macroscopic hysteresis loops. These findings expose the inertial delay of human behavioral adaptation—which we formalize as ``Fear Drift''—proving that large-scale contagion dynamics are  path-dependent and governed by gauge-mediated phase transitions.

The paper is organized as follows. In Section II, we formulate the general action of our model using the Doi-Peliti formalism, defining the complete stochastic action that couples the host compartmental dynamics, the pathogen gauge mediator, and the reactive immunity field. In Section III, we apply the Coleman-Weinberg mechanism to the immunity field, demonstrating how radiative loop corrections from the pathogen destabilize the naive vacuum and induce spontaneous symmetry breaking to generate a dynamical herd immunity threshold. In Section IV, we extract the classical saddle-point limit of the model to derive the macroscopic reaction-diffusion equations. This section is divided into the temporal dynamics, which reveals a non-linear, dressed Effective Reproductive Number ($R_{eff}$), and the spatial dynamics, which yields the thermodynamic free energy and the resulting macroscopic cross-diffusion fluxes. Finally, in Section V, we discuss the macroscopic signatures of this gauge-mediated contagion, explicitly mapping fundamental field-theoretic concepts—such as Debye screening, vacuum polarization, and the vanishing screening mass at the critical point—to observable epidemiological phenomena like behavioral spatial evacuation and critical opalescence. In section VI, we applied our formalism to the study case of Covid-19 in Germany.

\section{General Theoretical framework}
In this section, we shall define the action that will describe our model. Using  the Doi-Peliti formalism \cite{peliti1985path,doi1976second,tauber2014critical}, the following fields are introduced to represent Susceptibles ($\phi_S, \hat{\phi}_S$) and Infecteds ($\phi_I, \hat{\phi}_I$), and an auxiliary scalar field $\varphi(x, t)$ representing the pathogen concentration in the environment. To introduce Herd Immunity as a dynamic stability mechanism, we must introduce a new field that ``condenses'' to give the pathogen mass and blocking its propagation. This new field  $\Phi$ is called the Reactive Immunity Field (e.g., adaptive immune response density). At this step, it is fundamental to distinguish two type of herd inmunity: one called reactive immunity which is induced by the process from suceptibles to infected and then to recovered with inmunity and the proactive protection induced by vaccination. We shall now focus on the proactive herd inmunity introducing it as a field.  The action is now given as 
\begin{equation}
S = \int d^d x dt \left[ \mathcal{L}_{Hosts} + \mathcal{L}_{Pathogen} + \mathcal{L}_{Immune} + \mathcal{H}_{contagion} + \mathcal{H}_{recovery} - \mathcal{L}_{production} \right]
\end{equation}
where we have to define each part of the action:
\begin{enumerate}
    \item The Host Dynamics $\mathcal{L}_{Hosts}$: This term governs the baseline conservation and spatial mobility of the susceptible and infected populations before any interactions occur.
    \begin{equation}
        \mathcal{L}_{Hosts} = \sum_{a=S,I} \hat{\phi}_a (\partial_t - D_a \nabla^2) \phi_a
    \end{equation}
    \item The Pathogen Dynamics $\mathcal{L}_{Pathogen}$: This represents the inverse propagator of the pathogen in a vacuum, describing its natural environmental diffusion and biological half-life (bare mass $m_0^2$).
    \begin{equation}
        \mathcal{L}_{Pathogen} = \frac{1}{2} \hat{\varphi} (\partial_t - D_\varphi \nabla^2 + m_0^2) \varphi
    \end{equation}
    \item The Herd Immunity Field $\mathcal{L}_{Immune}$:It governs the dynamics of the reactive immunity field $\Phi(x,t)$. In the massless (scale-invariant) limit, it is defined by the Covariant Derivative, a self-interaction term (Social Resistance $\lambda$), and a bare mass term $\mu$:
   \begin{equation}
   \mathcal{L}_{Immune} = \underbrace{\hat{\Phi} \partial_t \Phi}_{\text{Time Evolution}} - \underbrace{D_\Phi \, \hat{\Phi} (\nabla - i g \varphi)^2 \Phi}_{\text{Gauged Diffusion}} 
   +\underbrace{\frac{\lambda}{4} (\hat{\Phi}^2 \Phi^2 - \hat{\Phi} \Phi)}_{\text{Self-Interaction}}+ \underbrace{\mu\hat{\Phi} \Phi}_{\text{mass term}}
   \label{inmune}
\end{equation}
where the $D_\Phi$ is the Immune Diffusion Coefficient and representing the mobility of the immune population. $\lambda$ can be called ``Social Resistance'' as the self-interaction is representing the natural tendency of immunity to wane or the ``cost'' of maintaining it. Let's recall that in the Doi-Peliti formalism, we are working in  a Euclidean Field Theory. It describes statistical weights $e^{-S}$ where the action $S$ represents the Hamiltonian/Free Energy functional. In the Doi-Peliti formalism, we map the state of the system $|n\rangle$ (n particles) to creation ($a^\dagger$) and annihilation ($a$) operators:
\begin{itemize}
    \item $a |n\rangle = \sqrt{n} |n-1\rangle$
    \item $a^\dagger |n\rangle = \sqrt{n+1} |n+1\rangle$
    \item Commutator: $[a, a^\dagger] = 1$
\end{itemize}
The operator representing the density of pairs is $a^\dagger a^\dagger a a$. We want to express this in terms of the number operator $\hat{n} = a^\dagger a$ (which corresponds to density $\phi^\dagger \phi$). Using the commutation relation $a a^\dagger = a^\dagger a + 1$:
\begin{equation}
a^\dagger a^\dagger a a = \hat{n}^2 - \hat{n}
\end{equation}
This explain why in eq(\ref{inmune}), in the self-interaction, we have to include a term $\lambda \hat{\Phi}\Phi$. 
So, one has:
\begin{equation}
    \mathcal{L}_{Immune} = \hat{\Phi} \partial_t \Phi - D_\Phi \hat{\Phi} (\nabla - i g \varphi)^2 \Phi + \frac{\lambda}{4} \hat{\Phi}^2 \Phi^2 + \underbrace{\left( \mu - \frac{\lambda}{4} \right)}_{\text{Net Mass Squared } (m_{eff}^2)} \hat{\Phi} \Phi
\end{equation}
    \item The Contagion Vertex $\mathcal{H}_{contagion}$ (Local Transmission): This operator describes the local transfer of state from Susceptible to Infected. A susceptible ($\phi_S$) is annihilated and an infected is created ($\hat{\phi}_I^2$ and $\hat{\phi}_S\hat{\phi}_I$) strictly in the presence of the pathogen field $\varphi$:
    \begin{equation}
        \mathcal{H}_{contagion} = \beta (\hat{\phi}_S \hat{\phi}_I - \hat{\phi}_I^2) \phi_S \varphi
    \end{equation}
    \item The Recovery Vertex $\mathcal{H}_{recovery}$:This describes the standard transition from Infected to Removed, pulling active infected individuals out of the dynamical Fock space:
    \begin{equation}
        \mathcal{H}_{recovery} = \gamma (\hat{\phi}_I - 1) \phi_I
    \end{equation}
    \item The Production Term $\mathcal{L}_{production}$: This linear coupling defines the infected population $I$ as the active source emitting the pathogen field, parameterized by the Epidemic Charge $\kappa$ (the viral shedding rate):
    \begin{equation}
        \mathcal{L}_{production} = \kappa \hat{\varphi} \phi_I
    \end{equation}
\end{enumerate}

\section{Coleman-Weinberg mechanism}
According the sign of $m_{eff}^2$ and $\lambda$, the herd inmunity can get a vacuum expectation value $\langle \Phi \rangle$ different from zero. In such a case, through the gauged-diffusion term, it will give a contribution to the pathogen mass proportional to $g^2\langle \Phi \rangle^2$ shortening the propagation length of the pathogen and limiting its propagation and in consequence the epidemic propagation. 

For the naive vacuum ($\Phi=0$) to be stable (representing a population that doesn't spontaneously become immune), the natural decay must overpower the quantum correction such that
\begin{equation}
    m_{eff}^2+\delta m^2 > 0
\end{equation}
where $\delta m^2$ are the loop corrections to the mass terms.

We shall assume that $m_{eff}$ is equal to zero such that our action is scale-invariant. Starting with a scale-invariant action describes the pre-pandemic state where the virus perceives the population as a continuous, infinite fuel source without barriers. It is why we shall also assume that $m_0$, pathogen mass, is also equal to zero which means that we are at a critical point ($R_0=1$) where a epidemic can start. We shall try to answer to the question if the loop corrections to our effective potential for the herd immunity $\Phi$ fields can induced a spontaneous symmetry breaking which will generate a finite propagation length for the pathogen and allowing to control the disease transmission.

In Quantum field theory, the standard Higgs mechanism with symmetry breaking (Landau-Ginzburg) is described putting at hand a mass term with the needed sign to generate the symmetry  breaking. In 1973, Coleman and Weinberg, in their paper\cite{coleman-weinberg}, demonstrated that spontaneous symmetry breaking can occur due to radiative (quantum loop) corrections even if the classical tree-level potential does not exhibit symmetry breaking.

So, this mechanism can also be applied to our action and effective potential  for the herd immunity field. To get the true vacuum structure, the Effective Potential $V_{eff}(\phi)$, which includes the energy contributions of all quantum fluctuations (loops) has to be computed.  To see if the Naive Vacuum ($\Phi=0$) is stable, we probe the system by imposing a constant background level of herd immunity, $\phi_c$. We then look at how the fluctuations of the virus and the population react to this background.We shift the field: $\Phi(x) \to \phi_c + \sigma(x)$.The masses of the fluctuations depend on this background level $\phi_c$:
\begin{itemize}
    \item Immune Mass ($m_\sigma^2$): $m_\sigma^2 = \frac{\lambda}{2} \phi_c^2$.
    \item Viral Mass ($m_\varphi^2$): $m_\varphi^2 = g^2 \phi_c^2$.The presence of background immunity $\phi_c$ gives the virus an effective mass (decay rate), limiting its range.
\end{itemize}
We integrate out the quantum fluctuations to find the effective potential $V_{eff}(\phi_c)$. This tells us the true energy density of the system.
\begin{equation}
    V_{1-loop} =\sum_{field}\frac{1}{2} \int \frac{d^4k}{(2\pi)^4} \text{Tr} \ln(k^2 + m_{field}^2(\phi_c))
\end{equation}
The total potential is then given as:
\begin{equation}
    V_{eff} = V_{tree} + V_{scalar-loop} + V_{viral-loop}
\end{equation}
The integral is divergent. We regulate it with a cutoff $\Lambda$ (e.g., the inverse size of a household, a neighboring or a city depending on the scale we are working) and introduce a renormalization scale $M$ (e.g., the total population density).We impose the ``Massless Renormalization Condition'':
\begin{equation}
    \frac{\partial^2 V_{eff}}{\partial \phi_c^2} \bigg|_{\phi_c=0} = 0
\end{equation}

It means that the ``Naive'' state ($\Phi=0$) remains a valid solution with zero mass locally. After calculating the integrals and applying the counter-terms,the effective potential is obtained assuming $g^4 \gg \lambda$ which means that the main contribution to the effective potential comes from the viral loops. One gets:
\begin{equation}
V_{eff}(\phi_c) = \frac{\lambda}{4!} \phi_c^4 + \left( \frac{3g^4}{64\pi^2} \right) \phi_c^4 \left[ \ln\left( \frac{\phi_c^2}{M^2} \right) - \frac{25}{6} \right]
\end{equation}

While the positivity of both coupling constants ($g^4, \lambda > 0$) guarantees global vacuum stability at large field values ($\phi \to \infty$), the emergence of a non-trivial vacuum requires a specific hierarchy at the renormalization scale: $\lambda \lesssim \mathcal{O}(g^4)$. This relation implies that for radiative symmetry breaking to occur, the Social Resistance ($\lambda$) must be sufficiently weak relative to the viral infectivity ($g$) to allow the radiative logarithmic corrections to generate a local minimum. If the social 'stiffness' is too high ($\lambda \gg g^4$), the potential barrier effectively suppresses the phase transition, locking the population in the naive susceptible state regardless of the pathogenic threat.

We find the new stable state by minimizing the potential ($\frac{\partial V}{\partial \phi_c} = 0$).This yields the emergent vacuum expectation value VEV (the herd immunity threshold):
\begin{equation}
    v = \langle \Phi \rangle = M \exp\left( \frac{11}{6} - \frac{4\pi^2 \lambda}{9 g^4} \right)
\end{equation}
The pathogen mass is then given by:
\begin{equation}
    m_{screen}^2 \propto D_\Phi \cdot g^2 \cdot v
\end{equation}
From this equation, one gets two regimes:
\begin{itemize}
    \item If infectivity ($g$) is Small: The exponent is a huge negative number, $v \approx 0$. The virus is too weak to trigger a massive immune response. We are in an endemic state. 
    \item If infectivity ($g$) is Large: The term $1/g^4$ becomes small. The exponent approaches a constant. $v$ becomes a significant fraction of $M$.The virus triggers a massive "condensation" of immunity that locks the system into a protected state. We get an emergent herd immunity.
\end{itemize}
This derivation proves that herd immunity is a first-order phase transition. Because of the logarithmic structure, there is a small ``bump'' (potential barrier) between $\phi=0$ and $\phi=v$.A small outbreak might not cross this barrier (it dies out, leaving the population naive). A large enough outbreak (or a variant with higher $g$) pushes the system over the barrier. The population ``rolls down'' to the new minimum $v$. Once at $v$, the system is ``stuck'' behind the barrier. Even if the virus disappears, the immunity remains (or wanes very slowly), acting as a permanent protection against re-infection.

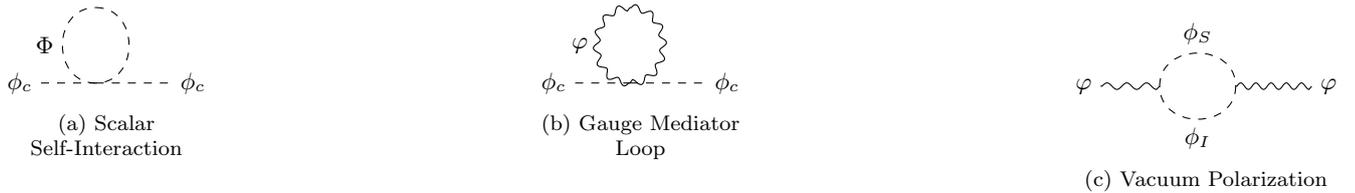
\begin{figure}[htpb]
    \centering
    \subfloat[Scalar Self-Interaction]{
        \begin{tikzpicture}[baseline=(a.base)]
            \begin{feynman}
                \vertex (a) {$\phi_c$};
                \vertex [right=1cm of a] (c);
                \vertex [right=1cm of c] (b) {$\phi_c$};
                \vertex [above=1cm of c] (t);
                \diagram* {
                    (a) -- [scalar] (c) -- [scalar] (b),
                    (c) -- [scalar, half left, edge label=$\Phi$] (t) -- [scalar, half left] (c),
                };
            \end{feynman}
        \end{tikzpicture}
    }
    \hfill
    \subfloat[Gauge Mediator Loop]{
        \begin{tikzpicture}[baseline=(a.base)]
            \begin{feynman}
                \vertex (a) {$\phi_c$};
                \vertex [right=1cm of a] (c);
                \vertex [right=1cm of c] (b) {$\phi_c$};
                \vertex [above=1cm of c] (t);
                \diagram* {
                    (a) -- [scalar] (c) -- [scalar] (b),
                    (c) -- [photon, half left, edge label=$\varphi$] (t) -- [photon, half left] (c),
                };
            \end{feynman}
        \end{tikzpicture}
    }
    \hfill
    \subfloat[Vacuum Polarization]{
        \begin{tikzpicture}[baseline=(a.base)]
            \begin{feynman}
                \vertex (a) {$\varphi$};
                \vertex [right=1cm of a] (v1);
                \vertex [right=1cm of v1] (v2);
                \vertex [right=1cm of v2] (b) {$\varphi$};
                \diagram* {
                    (a) -- [photon] (v1),
                    (v1) -- [scalar, half left, edge label=$\phi_S$] (v2),
                    (v2) -- [scalar, half left, edge label=$\phi_I$] (v1),
                    (v2) -- [photon] (b),
                };
            \end{feynman}
        \end{tikzpicture}
    }
    
    \caption{\textbf{Radiative Corrections in the QED-SIR Model.} (a) and (b) represent the 1-loop corrections driving the Coleman-Weinberg spontaneous symmetry breaking of the immunity field. (c) represents the vacuum polarization of the gauge mediator $\varphi$ by the host fields ($\phi_S, \phi_I$), which macroscopically manifests as the Debye screening penalty ($-S^3 I$) in the effective reproductive number.}
    \label{fig:feynman_loops}
\end{figure}
\section{ Classical limit of our model}

Our QED-inspired model with a Herd immunity field is not only able to reproduce the SIR equation models. In fact, one has two important terms in our action that are linking pathogen,Herd immunity, susceptibles and infectious field: 
\begin{eqnarray}
    L_{seagull}&=&g^2 \phi^2 \Phi^2 \\
    L_{production}&=& - \kappa \hat{\varphi}\phi_I
\end{eqnarray}
The first term is coming from the gauge interaction between herd immunity field and pathogen field. 

To get the macroscopic equation for our QED-inspired model, two distinct physical limits have to be taken into account:
\begin{itemize}
    \item The temporal evolution of the infection rates which is similar to non-conserved chemical reaction dynamics.
    \item the spatial transport dynamics which is implicit in our covariant derivative terms and diffusion coefficient. 
\end{itemize}

\subsection{Temporal dynamics}
In the Doi-Peliti formalism, taking the classical saddle-point limit ($\hat{\phi}_a = 1$) \cite{doi1976second,peliti1985path} reduces the operator equations to deterministic mean-field trajectories. The equation of motion for the pathogen field $\varphi$, sourced by the infected population $I$ with epidemic charge $g$, is:
\begin{equation}
    (\partial_t - D_\varphi \nabla^2 + m_0^2) \varphi + 2g^2 \Phi^2 \varphi = g I
\end{equation}
Because the environmental pathogen equilibrates much faster than the host population, the adiabatic limit ($\partial_t \varphi \approx 0, D_\varphi \approx 0$) can be applied. Substituting the complementary immune field $\Phi = N - S$ where $N$ is the total population, the  pathogen field becomes:
\begin{equation}
    \varphi \approx \frac{g}{m_0^2 + 2g^2(N-S)^2} I
\end{equation}
Assuming the interaction is a perturbative correction to the bare mass ($2g^2(N-S)^2 \ll m_0^2$), one gets:
\begin{equation}
    \varphi \approx \frac{\kappa}{m_0^2} I - \frac{2g^3}{m_0^4} \left( N^2 - 2NS + S^2 \right) I
\end{equation}
In the limit $g \rightarrow 0$, we recover the case studied in reference \cite{bernalalvarado2026}
Let's recall the equations obtained from our model to describe the susceptibles,$S(t)$,  and the infectious population,$I(t)$:
\begin{eqnarray}
    \frac{\partial S}{\partial t} &=& -\beta S \varphi \\
    \frac{\partial I}{\partial t} &=& \beta S \varphi - \gamma I \\
    \Rightarrow  \frac{\partial I}{\partial t} &=&  \frac{\beta \kappa}{m_0^2} S I - \frac{2\beta g^3}{m_0^4} \left( \underbrace{N^2 S I}_{\text{Term 1}} - \underbrace{2N S^2 I}_{\text{Term 2}} + \underbrace{S^3 I}_{\text{Term 3}} \right)-\gamma I
\end{eqnarray}
From a biological point of view, the $g$ and $\kappa$ parameters have to be related as their origin is from the same pathogen.  For simplicity, we shall assume $g=\kappa$:
\begin{equation}
    \frac{\partial I}{\partial t} = \frac{\beta g}{m_0^2} S I - \frac{2\beta g^3}{m_0^4} \left( \underbrace{N^2 S I}_{\text{Term 1}} - \underbrace{2N S^2 I}_{\text{Term 2}} + \underbrace{S^3 I}_{\text{Term 3}} \right) - \gamma I
\end{equation}
where the first term ($\frac{\beta g}{m_0^2} S I$) is the  classical Kermack-McKendrick contagion. The other terms are given by
\begin{itemize}
    \item Term 1 ($- \propto N^2 S I$) which can be called \textbf{Baseline Screening}. The total population size $N$ acts as a dielectric medium. Simply having a massive, dense population creates ``environmental friction'' that dilutes the per-capita exposure to the environmental pathogen field, suppressing the effective infection rate.
    \item Term 2 ($+ \propto 2N S^2 I$) which can be called \textbf{Density Clustering}. The positive sign indicates amplification. Because it scales with $S^2$, it captures the reality that susceptible people cluster together. This clustering creates local ``super-spreading'' hubs that artificially boost the infection rate above the mean-field average.
    \item Term 3 ($- \propto S^3 I$) which can be called \textbf{The Shielding Penalty}.  It is a severe, cubic negative feedback loop. If the susceptible density gets too high during an active outbreak, the mathematical penalty forces the infection rate to saturate and drop, mirroring the spontaneous behavioral isolation of a terrified population.
\end{itemize}
\subsubsection{Effective Reproductive Number}
In any compartmental model, the Effective Reproductive Number $R_{eff}$ is defined as the ratio of the total infection rate to the total removal/recovery rate ($\gamma I$).

\begin{eqnarray}
    \frac{\partial I}{\partial t} &= &\gamma \left( R_{eff}(S) - 1 \right) I \\
    R_{eff}(S) &=& \frac{\beta g}{\gamma m_0^2} S - \frac{2\beta g^3}{\gamma m_0^4} \left( N^2 S - 2N S^2 + S^3 \right)
\end{eqnarray}
The $R_{eff}$ can be rewrite as:
\begin{equation}
    R_{eff}(S) = \underbrace{R_0 \frac{S}{N}}_{\text{Classical } R_{eff}} - \underbrace{R_0 \frac{2 g^2}{m_0^2} \frac{S(N-S)^2}{N}}_{\text{The Shielding Penalty}}
\end{equation}
In the classical Kermack-McKendrick SIR model, the Effective Reproductive Number is strictly linear: $R_{classical} = R_0 (S/N)$. The epidemic only ends (crosses $R_{eff} = 1$) when the susceptible population is depleted purely by mass-action infection. This requires a massive number of infections to reach the classical Herd Immunity Threshold ($S_{HI}$).

In our gauge-mediated theory,  $R_{eff}$ is fundamentally non-linear. When the outbreak starts, $S \approx N$, so $(N-S)^2 \approx 0$. The shielding penalty is dormant. The virus spreads at the classical $R_0$ rate. As people get infected or isolated, the complement field $\Phi = (N-S)$ begins to grow. Because the penalty scales with $S(N-S)^2$, the $-S^3$ term heavily dominates the calculus. The sheer density of the population rearranging itself and isolating creates a massive mathematical friction against transmission. Because of this negative penalty, the $R_{eff}$ curve is ``squashed'' downward. It is forced to cross the critical threshold ($R_{eff} = 1$) at a much higher value of $S$ than the classical model predicts.

\subsubsection{Endemic case}
An endemic equilibrium occurs when the infected population stops growing and stops dying out \cite{hadeler1997backward,kribs2000simple,gomes2011epidemic}. Mathematically, this means the temporal derivative must be zero ($\frac{\partial I}{\partial t} = 0$) while the infected population remains non-zero ($I^* > 0$) \footnote{In a real-world scenario, reaching a true permanent endemic state requires a replenishment of susceptibles, such as waning immunity or new births, to balance the equation $\frac{\partial S}{\partial t} = 0$. We shall assume that this happened.}
This  means that 
\begin{equation}
    R_{eff}(S^*) = 1
\end{equation}
In the Classical SIR model, the effective reproductive number is a linear function of $S$, so one gets from the condition for endemic case:
\begin{equation}
    S^* = \frac{N}{R_0}
\end{equation}
This means classical epidemiology predicts exactly one possible endemic steady state. If a virus is endemic, it will always settle at this single predictable level.
Using the fractional density $s = S/N$ and the shielding coupling $\alpha = \frac{2g^2 N^2}{m_0^2}$, your endemic condition is now given by:
\begin{equation}
   R_0 s^* \left[ 1 - \alpha (1 - s^*)^2 \right] = 1 
\end{equation}
It is a cubic equation that can have uo to three real solutions. It means that for the exact same virus (same $R_0$, same $g$, same $m_0$), the population can settle into multiple different endemic states depending on how the pandemic started.

\subsection{The Spatial Dynamics}
While the reaction rate dictates state changes, the interaction also creates a thermodynamic pressure that drives spatial movement.
Spatial movement in a continuous macroscopic system is  governed by the gradients of its thermodynamic Free Energy.

When one goes from the microscopic Doi-Peliti action to the macroscopic classical limit, the spatial part of the Effective Action acts as a Landau-Ginzburg Free Energy functional \cite{tauber2014critical}. To find out how the population physically moves, we must use the standard rules of gradient flow for conserved densities, often called "Model B" dynamics in the Hohenberg-Halperin classification \cite{hohenberg1977theory}, which is mathematically equivalent to Cahn-Hilliard phase-separation kinetics \cite{cahn1958free}.


Integrating out the pathogen field generates an effective classical interaction potential (Free Energy) $V_{eff}$ for the host populations. The energy of the field $\varphi$ interacting with its source $gI$ is given by $E \approx -\frac{1}{2} \varphi (gI)$. Substituting the expanded $\varphi$:
\begin{equation}
    V_{eff} \approx - \frac{g^2 I^2}{2m_0^2} + \frac{g^4}{m_0^4} (N-S)^2 I^2
\end{equation}
Keeping in the free energy, only the terms depending on $S$, one gets
\begin{equation}
    f_{int} = C (N^2 - 2NS + S^2) I^2
\end{equation}
with $C = \frac{g^4}{m_0^4}$.

In thermodynamics, if one wants to know how a specific species (like the Susceptible population) will behave,   its Chemical Potential ($\mu_S$) has to be calculated \cite{degroot1984non}.The chemical potential is defined as the partial derivative of the Free Energy with respect to that species:
\begin{equation}
    \mu_S = \frac{\partial f_{int}}{\partial S} \propto 2 \left( \frac{g^4}{m_0^4} \right) S I^2
\end{equation}
Populations (like particles in a fluid) always flow from regions of high chemical potential (high stress/energy) to regions of low chemical potential (low stress/energy).The actual physical force pushing the population is the negative spatial gradient of the chemical potential:
\begin{equation}
    \mathbf{F}_{thermo} = - \nabla \mu_S
\end{equation}
With these elements, one can build the macroscopic flow of people. Fick's First Law of generalized diffusion \cite{onsager1931reciprocal,degroot1984non,kondepudi2014modern}states that the flux ($\mathbf{J}_S$) is equal to the mobility of the population ($\mathcal{M}$), multiplied by the amount of people available to move ($S$), multiplied by the driving force ($-\nabla \mu_S$)\cite{hohenberg1977theory,keller1971model}:

\begin{equation}
    \mathbf{J}_S = \mathcal{M} S (-\nabla \mu_S) = -\mathcal{M} S \nabla \left[ 2 \left( C (N^2 - 2NS + S^2) I^2 \right)  \right]
\end{equation}
It gives us three terms:
\begin{itemize}
    \item The Background Mass ($N^2 I^2$)$$\frac{\partial}{\partial S} \left( C N^2 I^2 \right) = 0$$ The total population $N$ is a uniform background constant. It creates a global "dielectric screening" that lowers the overall temporal infection rate, but because it doesn't depend on the local density of $S$, it exerts no spatial thermodynamic pressure.
    \item  The Linear Coupling ($-2NS I^2$)$$\frac{\partial}{\partial S} \left( -2 C N S I^2 \right) = -2 C N I^2$$The force is $-\nabla \mu_S = \nabla (2 C N I^2) = 4 C N I \nabla I$. When multiplied by the mobility $\mathcal{M}S$, this creates a flux $\mathbf{J} \propto N S I \nabla I$. This is the Standard Linear Fear Drift. It represents susceptible people acting as isolated individuals, looking strictly at the pathogen gradient ($\nabla I$) and running away. This is identical to classic Keller-Segel chemotaxis equations.
    \item The Non-Linear Repulsion ($S^2 I^2$)$$\frac{\partial}{\partial S} \left( C S^2 I^2 \right) = 2 C S I^2$$ The force is $-\nabla \mu_S = -\nabla (2 C S I^2) = -2 C I^2 \nabla S - 4 C S I \nabla I$. 
    It generates the Infection-Activated Dispersion (fleeing the crowd) and the Non-Linear Fear Drift (panic compounding as density increases).
\end{itemize}
Multiplying by the mobility ($\mathcal{M} S$), one gets the complete, rigorously derived spatial flux:
\begin{equation}
    \mathbf{J}_S = \underbrace{- 2 C \mathcal{M} S I^2 \nabla S}_{\text{Dispersion}} + \underbrace{4 C \mathcal{M} S (N - S) I \nabla I}_{\text{Modulated Fear Drift}}
\end{equation}

\subsection{Complete Macroscopic Equations}
Putting all previous results together, one gets the complete macroscopic equations for $S$ and $I$:
\begin{equation}
    \partial_t S = - \nabla \cdot \mathbf{J}_{total} - \left[ \beta_0 S I - \frac{2\beta_0 g^2}{m_0^2} (N - S)^2 S I \right]
\end{equation}
\begin{eqnarray}
    \frac{\partial S}{\partial t} &=&  \underbrace{D_S \nabla^2 S}_{\text{Normal Diffusion}} + \underbrace{\nabla \cdot \left( \chi_0 S I^2 \nabla S \right)}_{\text{Dispersion}} - \underbrace{\nabla \cdot \left[ 2 \chi_0 S (N-S) I \nabla I \right]}_{\text{Modulated Fear Drift}} \nonumber \\
    &&- \underbrace{\left[ \beta_0 S I - \frac{2\beta_0 g^2}{m_0^2} (N - S)^2 S I \right]}_{\text{Dressed Reaction Rate}}\\
  \frac{\partial I}{\partial t} &=& \underbrace{D_I \nabla^2 I}_{\text{Normal Diffusion}} + \underbrace{\left[ \beta_0 S I - \frac{2\beta_0 g^2}{m_0^2} (N - S)^2 S I \right]}_{\text{Dressed Reaction Rate}} - \underbrace{\gamma I}_{\text{Recovery}}
\end{eqnarray}
where $\beta_0 = \frac{\beta g}{m_0^2}$ is the bare classical transmission rate and $\chi_0 = 2 \mathcal{M} \frac{g^4}{m_0^4}$. $\mathcal{M}$ (Host Mobility)is the baseline kinematic mobility or diffusion capacity of the susceptible population. If people are physically unable to move (e.g., due to a strict, physically enforced lockdown), $\mathcal{M} \to 0$, and the behavioral spatial recoil ($\chi_0$) collapses. The first equation is the continuity equation for $S(t)$.

\section{Discussion: Macroscopic Signatures of Gauge-Mediated Contagion}
The derivation of the unified macroscopic PDEs establishes a relation  between the stochastic Doi-Peliti field theory and  epidemiological observables. By extracting the classical mean-field limit of the gauge-mediated interaction after the integration on the pathogen field,  non-local and non-linear transmission phenomena appear. 

In the QED-inspired formalism, Debye Screening is the process by which the susceptible population absorbs and shields the pathogen field, impeding it from reaching distant hosts. The pathogen's effective range is related to the inverse  screening length $\lambda_D = \frac{1}{m_0\sqrt{1-R_0}}$. 
Our macroscopic PDE reveals how the screening due to Debye mass has be understood in terms of macroscopic behavior.   The conserved spatial transport of the susceptible population is governed by the non-linear ``Fear Drift'' and ``Infection-Activated Dispersion'':
\begin{equation}
    \mathbf{J}_{screen} = - 2 \chi_0 S^2 I \nabla I - \chi_0 S I^2 \nabla S
\end{equation}
This flux is the classical, fluid-dynamic realization of a screening cloud. The susceptible population  does not remain static; it physically redistributes itself away from the pathogen gradient ($\nabla I$) and dilutes its own local density ($\nabla S$). This spatial rearrangement actively shields the surrounding environment, acting as the macroscopic analog to the diamagnetic expulsion of a gauge field. It is this exact behavioral evacuation that generates the finite interaction range $\lambda_D$ observed in the pathogen propagator.

As shown in our previous work \cite{bernalalvarado2026}, when a pathogen moves through a vacuum of susceptibles, it constantly couples to the hosts, modifying its own propagation by dressing its mass via the Vacuum Polarization $\Pi(q)$. 
In our derived temporal reaction rate, the classical mass-action contagion ($\beta_0 S I$) is modified by a polynomial series:
\begin{equation}
    \frac{\partial I}{\partial t}\bigg|_{reaction} \propto \beta_0 S I - \frac{2\beta_0 g^2}{m_0^2} (N^2 S I - 2N S^2 I + S^3 I)
\end{equation}
This non-linear saturation is the macroscopic manifestation of the dressed interaction vertex. The cubic term ($-S^3 I$) mathematically proves that as the density of the susceptible vacuum increases, the probability of infection does not scale linearly. Instead, the dense host medium dynamically suppresses the effective transmission, $R_{eff}$. This density-dependent suppression is the classical equivalent of the 1-loop self-energy correction, where the ``dielectric constant'' of the epidemiological vacuum heavily penalizes transmission in overcrowded states.

In previous work \cite{bernalalvarado2026}, we have shown how the QED-inspired model  is able  to predict the onset of a pandemic through the lens of a  phase transition. The well-known epidemic threshold is re-formulated as a symmetry-breaking event where the renormalized mass vanishes ($m_R^2 \to 0$). As the system approaches this critical threshold ($R_0 \to 1$), the Debye screening length diverges ($\lambda_D \to \infty$). In our macroscopic equations, this limit occurs when the stabilizing background mass term ($N^2 S I$) is  balanced by the linear coupling amplification ($2N S^2 I$). When this effective mass vanishes, the restoring forces of the system are neutralized, and the screening fails. At this precise moment, the spatial ``Fear Flux''  and the temporal non-linearities are competing. Because the screening length is infinite, local fluctuations in the pathogen density are no longer damped. The population is torn between fleeing (dispersion) and clustering, resulting in scale-free, macroscopic density fluctuations. In epidemiological data, this manifests as Critical Opalescence: a sudden, dramatic increase in the variance of case counts and the emergence of self-similar infection clusters across multiple spatial scales.

\section{Empirical Application: Macroscopic Dynamics of COVID-19 in Germany}
\label{sec:empirical_application}

To test the macroscopic validity of our QED-inspired epidemiological model, we applied the formalism to the spatiotemporal evolution of the SARS-CoV-2 pathogen in Germany.  We performed a mean-field approximation over 400 spatial districts (AGS codes), representing a federal population of approximately 83 million. By treating the total active infected population $I(t)$ as an extensive thermodynamic variable (spatially summed) and the effective screening proxy $m_{eff}(t)$ as an intensive gauge field (spatially averaged), we mapped the empirical phase space of the pathogen-host interaction. 

This coarse-graining method filters out microscopic stochastic noise, isolating the  thermodynamic forces that govern the system.  Distinct topological signatures that cannot be adequately explained by classical, memoryless mass-action models (e.g., standard SIR differential equations) are obtained.

\subsection{Data Acquisition and Spatiotemporal Preprocessing}
\label{subsec:data_sources}

The  validation of our theoretical model is based  on two  high-resolution data streams capturing both the kinematic state of the epidemic and the behavioral gauge field across Germany. The dataset is covering the temporal window from the onset of the pandemic through the post-vaccination endemic phase (March 2020 to early 2023), resolved at the spatial granularity of over 400 German districts (\textit{Landkreise} and \textit{kreisfreie Städte}, identified by their AGS codes).

First, the macroscopic state variable—the daily active infected population $I(x,t)$—was constructed using official epidemiological surveillance data provided by the Robert Koch Institute (RKI) \cite{rki_covid_data}. The raw daily incidence data was processed to remove administrative reporting anomalies (such as the \textit{Weihnachtseffekt} during winter holidays and weekend reporting lags) using a Gaussian filter and temporal coarse-graining, ensuring the thermodynamic state variable reflected true transmission dynamics rather than statistical artifacts.

To preserve the physical dimensions of the variables during spatial upscaling to the federal level, the extensive incidence variable $I(t)$ was spatially summed, while the intensive gauge field $m_{eff}(t)$ was spatially averaged. This  thermodynamic aggregation allowed us to link microscopic, district-level interactions with the macroscopic, federal-level phase transitions analyzed in this study.
\subsection{Empirical Extraction of the Screening Mass via Spatial Correlations}
\label{subsec:mass_extraction}

In our gauge model, the behavioral and immunological interventions of the host population act as a screening field that dampens the spatial propagation of the pathogen. The strength of this screening is quantified by the effective mass parameter $m_R$ (or $m_{eff}$). To empirically extract this time-dependent and spatially heterogeneous gauge field parameter from the data, we analyzed the spatial decay of incidence correlations between German districts.

We isolated specific epidemic windows corresponding to the Omicron wave from December 2021 to April 2022 and constructed daily incidence time series $I_i(t)$ for each district $i$, mapped via their AGS (Allgemeiner Gemeindeschlüssel) codes. For every pair of districts $(i, j)$, we computed the normalized Pearson cross-correlation of their incidence time series over the observation window $T$:
\begin{equation}
    r_{ij} = \frac{1}{T} \sum_{t=1}^{T} \tilde{I}_i(t) \tilde{I}_j(t)
\end{equation}
where $\tilde{I}(t)$ is defined as the standardized incidence (zero mean, unit variance). Simultaneously, we computed the pairwise great-circle distance matrix $D_{ij}$ using the geographic centroids of the districts. 

To  extract the correlation length, the pairwise correlations $r_{ij}$ were binned by geographic distance (e.g., in 20 km intervals up to 500 km). For each district $i$, we fitted the binned empirical correlation curve using an exponential screening kernel:
\begin{equation}
    C(d) = A_i e^{-d/\xi_i} + C_i
\end{equation}
where $\xi_i$ is the empirical spatial correlation length, $C_i$ accounts for a baseline systemic background correlation and $d$ is the distance to the district center.

\subsubsection{Justification of the Exponential Propagator vs. Yukawa Potential}

In a continuous, infinite 3D medium, a massive scalar field propagates according to the Yukawa potential, $V(d) \propto e^{-md}/d$ \cite{bernalalvarado2026}. However, applying the exact Yukawa kernel to discrete epidemiological data introduces structural instabilities. 

At district-level data, short-distance measurements are highly sensitive to centroid approximation errors and local demographic heterogeneities, which distort the fit. Second, human mobility within a finite country with national borders does not emulate an infinite homogeneous medium, obscuring the clean asymptotic $1/d$ behavior. Attempting global fits with the explicit $1/d$ geometric factor resulted in overfitting the initial distance bins, thereby destabilizing the estimation of the correlation length $\xi$.

Consequently, we opted for the  form $e^{-d/\xi}$. This function isolates the far-field exponential screening—which dominates the macroscopic topological decay—while mathematically bypassing the short-distance singularities. Within this gauge-mediated field theory, the screening mass $m_{R,i}$ for each district is then given by:
\begin{equation}
    m_{R,i} = \frac{1}{\xi_i}
\end{equation}
This procedure yields a  data-driven proxy for the effective mass (measured in km$^{-1}$). 

\subsection{Empirical Proxy for the Gauge Field: Google Mobility Data}
\label{subsec:mobility_data}
To apply our model, the movility of the population has to be parametrized. For that,we utilized the publicly available Google COVID-19 Community Mobility Reports \cite{google_mobility_data}. This dataset provides high-resolution, anonymized, and aggregated mobility trends across various location categories (e.g., transit stations, workplaces, retail, and recreation) relative to a pre-pandemic baseline.

Within our theoretical model, macroscopic reductions in population mobility are mathematically isomorphic to an increase in the environmental screening mass of the pathogen. By aggregating these mobility indices at the district (AGS) and subsequently federal levels, we derived a continuous, data-driven proxy for the inverse environmental interaction time, $m_{eff} \sim \tau_{env}^{-1}$. 

\subsection{Sliding Gauge Dynamics and the Effective Reproduction Number ($R_{eff}$)}
\label{subsec:reff_analysis}

A standard metric for epidemic growth is the empirical effective reproduction number, $R_{eff}^{emp}(t) = 1 + \frac{1}{\gamma I(t)} \frac{dI(t)}{dt}$. Traditional compartmental models have difficulties to fit this continuous metric without imposing arbitrary parameter jumps or artificial forcing functions to account for behavioral changes. 

In our model, behavioral and immunological interventions are naturally absorbed into the effective pathogen mass through the gauge coupling. To demonstrate this, we implemented a Gauge-Sliding optimization algorithm over the  data. 
A sliding temporal window is iteratively solved for the pathogen's bare transmission rate $\beta$ while continuously incorporating the empirical screening mass $m_{eff}(t)$ and the immune condensate field $\Phi(t)$.

The theoretical reproduction number derived from the dressed propagator is given by:
\begin{equation}
    R_{eff}^{Gauge}(t) = \frac{\mathcal{M}(t) R_0 \frac{S(t)}{N}}{1 + \alpha \Phi_{eff}(t)^2}
\end{equation}
where $\mathcal{M}(t)$ modulates the local mobility. The sliding window analysis demonstrated that the QED-inspired $R_{eff}^{Gauge}(t)$ better explained the empirical $R_{eff}^{emp}(t)$ extracted from smoothed Robert Koch Institute (RKI) data much better than SIR models.
\begin{figure}
    \centering
    \includegraphics[width=1\linewidth]{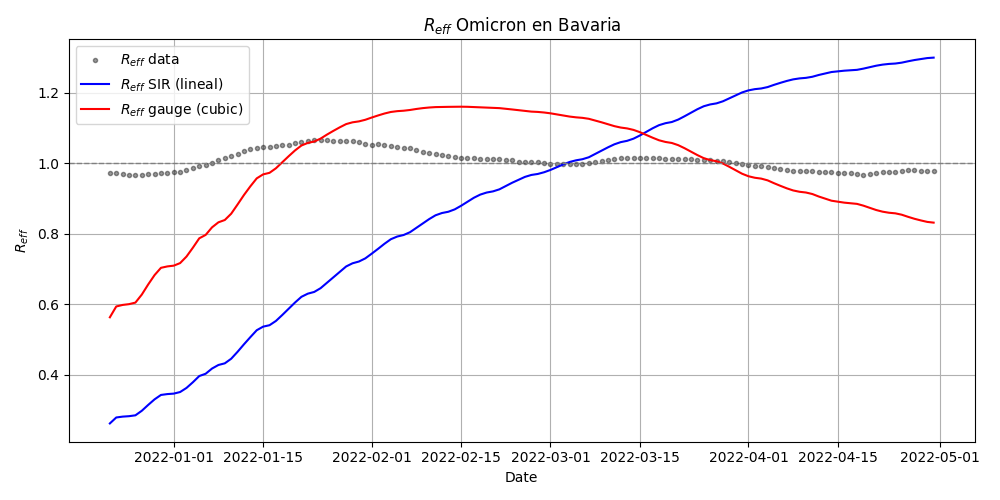}
    \caption{\textbf{Comparative Dynamics of Gauge-Mediated vs. Classical SIR Reproduction Numbers.} The black dots represent the $R_{eff}$ computed directly from the data. The QED-inspired gauge model ($R_{eff}^{Gauge}$, red curve) dynamically incorporates the behavioral shielding mass $m_{eff}(t)$. In contrast, the classical mass-action model ($R_{eff}^{SIR}$, blue curve) relies  on the depletion of the susceptible population, resulting in a structural temporal lag and an inability to reproduce the sharp suppression of transmission without arbitrary parameter forcing.}
    \label{fig:reff_comparison}
\end{figure}

\subsection{Spontaneous Symmetry Breaking and Endemic Bistability}
\label{subsec:bistability}

To investigate the static topology of the system, we constructed the empirical potential landscape, defined as $V(I) = -\ln P(I)$, where $P(I)$ is the stationary probability distribution of the incidence during the post-vaccination endemic phase (August 2021 - January 2023). This landscape acts as the epidemiological equivalent of a Landau free-energy potential.

The empirical data reveals a  first-order phase transition signature governed by the control parameter $m_{eff}$ (Fig. \ref{fig:phase_space}A). For low values of the screening field ($m_{eff} \leq \text{median}$), the system relaxes into a single, deep potential well corresponding to a high-incidence vacuum state. In this regime, the absence of strong behavioral or immunological shielding ensures that sustained transmission is the only stable equilibrium. 

However, as the effective mass surpasses the median threshold, the potential undergoes a radical topological deformation, exhibiting \textit{spontaneous symmetry breaking}. The landscape develops a pronounced double-well structure, revealing two distinct local minima separated by a potential barrier. This mathematical feature empirically proves \textit{macroscopic bistability}: under the exact same stringent level of societal shielding, the national population can stabilize into two mutually exclusive metastable states—either successful suppression (low incidence) or sustained endemia (high incidence). This result  validates the cubic nonlinearities (interaction vertices) introduced by the gauge-mediated spatial shielding in our effective Lagrangian.

\subsection{Phase Space Trajectories, Hysteresis, and "Fear Drift"}
\label{subsec:hysteresis_fear_drift}

\begin{figure}[htbp]
    \centering
    \includegraphics[width=0.95\textwidth]{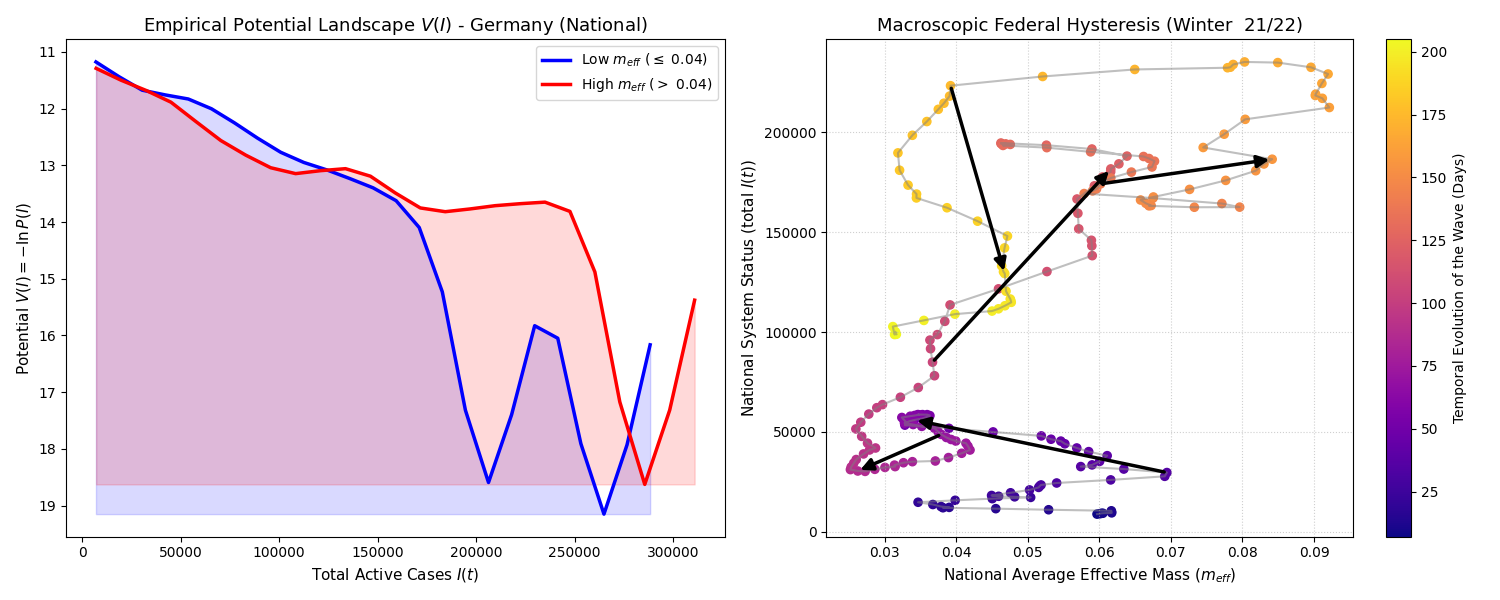} 
    \caption{\textbf{Macroscopic Signatures of Gauge-Mediated Phase Transitions.} (\textbf{A}) Empirical potential landscape $V(I) = -\ln P(I)$. A strong $m_{eff}$ (blue curve) induces spontaneous symmetry breaking, generating a double-well potential indicative of endemic bistability. (\textbf{B}) Phase space trajectory of the winter wave in the $(m_{eff}, I)$ plane. The macroscopic hysteresis loop demonstrates path dependence and the inertial delay of the  gauge field (Fear Drift).}
    \label{fig:phase_space}
\end{figure}

While the potential landscape defines the static equilibria, the kinematic evolution of the epidemic reveals the non-Markovian nature of the dynamics. Mapping the trajectory of the 2021/2022 Omicron winter wave in the $(m_{eff}, I)$ phase space exposes a massive, counter-clockwise \textit{hysteresis loop} (Fig. \ref{fig:phase_space}B).

In classical memoryless models, trajectories are generally reversible. The open loop observed in our data is the visual signature of \textit{path dependence}, capturing the inertial delay of the social gauge field—a phenomenon we term the \textbf{"Fear Drift"}. The three distinct thermodynamic phases of the histeresis cycle are given by:

\begin{enumerate}
    \item \textbf{The Inertial Lag (Exponential Surge):} During the wave's onset, the incidence $I(t)$ surges exponentially before the societal effective mass $m_{eff}$ can react, due to reporting delays and political hesitation. The system travels up the left branch of the phase space, experiencing maximum growth unhindered by the gauge field.
    
    \item \textbf{The Braking Phase (Peak Shielding):} As the "fear" signal propagates, the social gauge field couples strongly with the pathogen. The effective mass $m_{eff}$ is driven to its maximum value, forcing the system over the potential barrier and halting the epidemic growth. This cycle represents the dissipative social and economic cost induced by the delayed response.
    
    \item \textbf{Thermodynamic Collapse and Relaxation:} At the closing phase of the cycle, the incidence  decreases even if  $m_{eff}$ relaxes. In a linear model, relaxing shielding would trigger an immediate rebound. In our gauge model, the massive wave depletes the macroscopic susceptible field $S(t)$ (condensate depletion), destroying the interaction vertex. The wave collapses thermodynamically due to a lack of ``reaction fuel.'' Consequently, the societal shielding mass $m_{eff}$ decreases as a delayed, inertial response to the already collapsing incidence, safely returning the social gauge field to its baseline coordinates without inducing a secondary surge.
\end{enumerate}

Collectively, these empirical signatures—gauge-driven $R_{eff}$ modulation, bistability, and hysteresis—demonstrate that pathogen propagation in large populations behaves not as a simple kinetic reaction, but as a  physical system governed by gauge-mediated phase transitions.

\section{Conclusion}
In this paper, we have demonstrated that the non-local characteristics and non-linear behavioral dynamics of epidemics emerge  from a single, gauge-mediated interaction. By treating the pathogen as a mediator field coupled to the host population, we  relate the microscopic stochastic field theory with deterministic, macroscopic population dynamics.

Our main result is that the Coleman-Weinberg radiative symmetry breaking and the macroscopic ``Fear Flux'' are mutually dependent manifestations of the exact same physical process. In the microscopic regime, the radiative loop corrections destabilize the naive epidemic vacuum, generating a stable vacuum expectation value for the Reactive Immunity Field. Macroscopically, integrating out the gauge mediator yields an effective thermodynamic Free Energy that depends on $S^2$. The gradient of this Free Energy drives a conserved spatial transport characterized by a Modulated Fear Drift ($\mathbf{J} \propto S(N-S)I\nabla I$). This proves that the phase transition into the new immune vacuum is executed physically by the susceptible population redistributing itself to avoid the pathogen gradient, thereby generating the macroscopic screening cloud that cages the disease.

Furthermore, the classical limit of our temporal equations shows  that this gauge-mediated interaction dresses the standard contagion vertex. The  effective reproductive number, $R_{eff}$, features a severe cubic shielding penalty ($-S^3 I$) that forces the transmission rate to drop  earlier than predicted by classical SIR models. This QED-inspired penalty not only accounts for premature herd immunity, but its cubic nature proves the existence of endemic bistability. Depending on the trajectory of the population's spatial shielding, a highly infectious pathogen can exhibit hysteresis, locking into either a high-transmission or low-transmission steady state.

The empirical application of our formalism to the German COVID-19 dataset provides a macroscopic validation of the gauge-theoretic approach to epidemiology. By coarse-graining the spatiotemporal data into federal thermodynamic observables, we successfully extracted the physical signatures of our proposed Lagrangian. The empirical emergence of a double-well potential confirms that strict behavioral shielding induces topological bistability, allowing the system to stabilize in mutually exclusive endemic or suppressed states under identical macroscopic conditions. Furthermore, the massive hysteresis loop observed during the Omicron winter wave isolates the ``Fear Drift''—the societal gauge inertia that decouples the pathogen's exponential surge from the delayed behavioral response. This path-dependent trajectory demonstrates that the epidemic phase space is not reversible;
Finally, these results  establish that human contagion  is a complex, interacting field theory where the psychological and behavioral effects described through our gauge field dictate the macroscopic topology of the disease.
\begin{acknowledgments}
We acknowledge financial support from SECIHTI, SNII, and U. of Gto. (M\'exico).
\end{acknowledgments}


%

\end{document}